\documentclass[conference]{IEEEtran}
\usepackage{amsmath,amsfonts}
\usepackage{algorithmic}
\usepackage{algorithm}
\usepackage{array}
\usepackage[caption=false,font=footnotesize,labelfont=rm,textfont=rm]{subfig}
\usepackage{textcomp}
\usepackage{stfloats}
\usepackage{url}
\usepackage{verbatim}
\usepackage{graphicx}
\usepackage{cite}
\usepackage{multirow}
\usepackage{arydshln}
\usepackage{marvosym}

\def\BibTeX{{\rm B\kern-.05em{\sc i\kern-.025em b}\kern-.08em
    T\kern-.1667em\lower.7ex\hbox{E}\kern-.125emX}}
    
\begin{document}

\title{LLM: LSTM Look-Ahead Moving Target Defense Based on Historical Malicious Scan}


\author{\IEEEauthorblockN{Yu Li}
}

\maketitle

\begin{abstract}
Network scanning is a critical preliminary step for most adversaries to gain essential information before launching cyber attacks. Moving Target Defense (MTD) based on IP shuffling has emerged as a proactive defense strategy to counteract these reconnaissance efforts.  Unlike static, reactive defense techniques, IP shuffling introduces randomness by dynamically reassigning network addresses, making it more challenging for attackers to identify and track targets. However, current IP shuffling methods face three key challenges: 1) limited scalability across different network topologies, 2) inherent reconfiguration overhead even in the absence of an active attack, and 3) the need for large-scale unused address blocks. To address these issues, we propose LSTM Look-ahead Moving Target Defense (LLM). Our approach is the first attempt using a Long Short-Term Memory (LSTM) network to predict future target addresses that attackers will likely scan. Ensemble learning is used to improve robustness to different scanning behaviors. We introduce a dynamic mutation mechanism to enhance adaptability. Compared to the baseline mutation strategy, LLM performs better in both security and overhead.

\end{abstract}


\section{Introduction}
Cyber attacks typically begin with target selection and information gathering. A concealed attacker conducts internal reconnaissance, using scanning techniques to identify potential targets. This process of reconnaissance can extend over a significant period, typical example is Advanced Persistent Threat (APT). Unfortunately, the static nature of many networks aids attackers in acquiring critical information. Despite the presence of defense mechanisms like firewalls and intrusion detection systems, these defenses are often passive and static, allowing attackers to detect and bypass them with relative ease.

Moving Target Defense (MTD) is a transformative technique that shifts networks from static to dynamic configurations. MTD continuously alters the configuration of a network, making it unpredictable and more resilient to network reconnaissance. One such classical MTD approach is IP shuffling (or host address mutation), which disrupts network scans by periodically changing the IP addresses of hosts. Specifically, the IP that the host used is virtual IP (vIP), assigned from available address pools. At regular intervals, the vIP undergoes random and unpredictable shuffling, confusing and interrupting attacker reconnaissance.

However, several limitations of current MTD based on IP shuffling need to be addressed. First, the scalability of mainstream IP reshuffling methods is limited and not suitable for dynamic networks. When the network topology changes, it takes extra time to calculate the available address space and Wait for the routing table to convergence. Second, the mutation frequency is statically determined by the system, even when no reconnaissance is being conducted by adversaries. This leads to unnecessary resource consumption, as the MTD-protected system continues to use computational and communication resources to maintain a high mutation frequency. Third, the mutation method is most effective with Class A private addresses due to their large number of unallocated addresses. The mutation space typically requires at least $2^5$ available addresses to provide fast and unpredictable service for each host \cite{DBLP:journals/tifs/JafarianAD15}. 

In this paper, we propose LSTM Look-ahead MTD (LLM) to address the limitations discussed earlier. Unlike existing solutions, LLM focuses on fixed subnets where the number of available addresses is limited (e.g., a typical subnet with a mask of 255.255.255.0). This means that no changes are required to the traditional network, thereby avoiding the overhead of rerouting. At the same time, there is no need to consider any complex address block allocation algorithms, as the available address blocks are limited and fixed. To minimize the overhead of address mutation, LLM adopts a hybrid mutation strategy, selecting unused address ranges based on deep learning predictions and adapting dynamically according to scan frequency. We deploy an IP shuffling strategy using the smallest possible address space, which requires us to carefully select those vIPs that are least likely to be scanned by an attacker in the following period of time. We employ Long Short-Term Memory (LSTM) networks to learn the historical scanning behaviors and predict future scanning addresses. The prediction results use ensemble learning to enhance the accuracy and robustness. All vIPs will bounce around in a limited address space, always dodging those risky addresses that an attacker is about to scan. Additionally, we use Software-Defined Networking (SDN) as our experimental platform due to its flexibility in network management. We evaluate LLM against various reconnaissance behaviors and demonstrate its effectiveness in thwarting scanning attacks.

\section{Related Work}\label{sec:Relate Works}
The concept of MTD was first introduced at the U.S. National Cyber Leap Year Summit in 2009. MTD aims to thwart attacks by creating dynamic environments that continuously shift and evolve over time. By altering the properties of the target system, on which the attack depends, MTD significantly increases the complexity and cost for adversaries. Among diverse MTD approaches, we focus on host address mutation at the network level. Several works have proposed IP address randomization to thwart adversarial reconnaissance. The DyNAT system \cite{DBLP:journals/ijcritis/Michalski06} is the first to use network address translation (NAT) to deploy dynamic IP addresses within networks. However, these approaches necessitate the disruption of network services and the deployment of additional devices. Network Address Space Randomization (NASR) was proposed in \cite{DBLP:journals/cn/AntonatosAMA07} to periodically shuffle IP addresses based on the Dynamic Host Configuration Protocol (DHCP). While NASR reduces the predictability of IP addresses, it can disrupt active connections during the reassignment process. Furthermore, due to the limitations imposed by IP address lease times, the maximum mutation frequency is once every 15 minutes.

Lin et al. \cite{DBLP:conf/srds/LinWHWC16} optimize the performance-security trade-off by dynamically controlling the mutation interval using a genetic algorithm. The mutation rate adapts based on the scanning speed of adversaries, enabling responsive adjustments. However, this work focuses on large-scale networks (over 1,000 hosts and equivalent address blocks), rather than small networks. Random Port and Address Hopping (RPAH) \cite{DBLP:conf/trustcom/LuoWWHCS15} and Flexible Random Virtual IP Multiplexing (FRVM) \cite{DBLP:conf/trustcom/SharmaK0LCM18} enhance the security of RHM by combining IP address mutation with host port changes, thus expanding the mutation space and defending against port scanning. While these approaches increase the available address space for address-constrained networks, they lack adaptivity in their services.

Jafarian et al. \cite{DBLP:conf/sigcomm/JafarianAD12} introduced Random Host Mutation (RHM), a novel mutation strategy that does not alter the actual IP address. Additionally, several works have enhanced the security and performance of RHM. Intelligence-Driven Host Address Mutation (ID-HAM) \cite{DBLP:journals/tifs/ZhangXSKG23} uses deep reinforcement learning techniques to improve RHM, alleviating network resource waste caused by inefficient mappings between mutated IP addresses and hosts. These efforts aim to enhance the adaptability and security of RHM while reducing performance overhead. However, they incorporate satisfiability modulo theory to calculate available mutation actions, which can be computationally expensive, particularly during network topology changes. 

Inspired by address prediction attacks against MTD \cite{DBLP:journals/tifs/AlmohriW020, DBLP:journals/tdsc/ManiHBMKKVWJAY23}, our work aims to predict future attacker scanning behaviors based on known address sequences. Address prediction not only reduces the overhead associated with attack surface movement but also integrates more effectively with cyber deception. Specifically, we can obscure real hosts by ensuring that the IP addresses attackers scan are always those of honeypots or honeynets. To the best of our knowledge, our work is the first to design an RNN-based prediction model for MTD to defend against reconnaissance in networks with dynamic topologies.

\section{Threat Model}\label{sec:Models}
\begin{figure}[htbp]
\centerline{
\includegraphics[width=0.9\linewidth, trim=10 30 10 40]{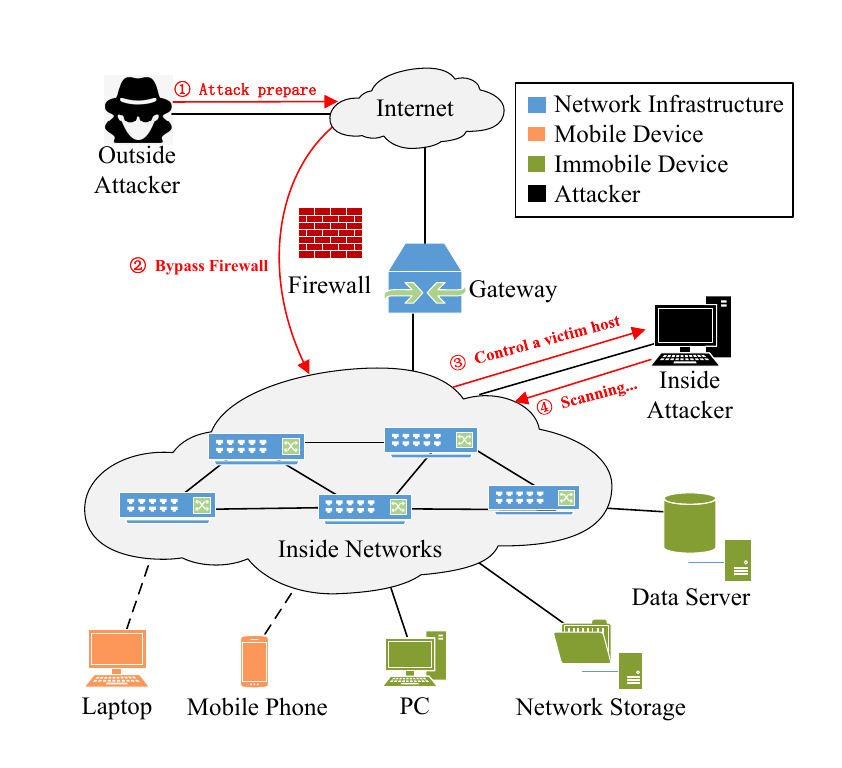}}
\caption{\centering{Threat model}}
\label{fig:fig1}
\end{figure}

Fig. \ref{fig:fig1} illustrates the threat model, focusing on the scenario where an internal adversary engages in network reconnaissance. We assume that all these devices are connected within a small Local Area Network (LAN) and an internal intruder can initially control at least one device, enabling them to send probe packets. We do not focus on external attackers, as most gateways are assigned dynamic IP addresses by network service providers due to the scarcity of available IPv4 public addresses \cite{DBLP:journals/ton/WangWWDXLS20}. Most external scanning is generally easier to detect and block via firewalls. After exploiting zero-day vulnerabilities to intrude into the network, the internal attacker scans the entire LAN stealthily and persistently. 

Reconnaissance plays a critical role throughout the cyber kill chain. According to \cite{DBLP:journals/csur/RoySAKL23}, reconnaissance can be categorized into two types: external reconnaissance (e.g., social engineering techniques) and internal reconnaissance (e.g., network scanning techniques). This work focuses on the internal attacker, whose goal is to gather network-level information through scanning. Specifically, we are concerned with risky behaviors involving probes sent to any available address within the local area network. Our primary focus is on host discovery, which represents the first step of network scanning. Scanning behavior can be characterized by three key indicators: range, speed, and strategy.

\subsubsection{Scanning Range} The scanning range refers to the address blocks that attackers deem likely to contain valuable hosts. It is commonly assumed that attackers can determine the complete address range of an internal network through preparatory activities such as network sniffing and listening. Therefore, the defender's goal is to use maximum entropy for unpredictable allocation of vIP within a determined address space, thereby increasing the time required for attackers to scan the network.

\subsubsection{Scanning Speed} Scanning speed is a critical indicator of malicious scanning behavior. A large number of ARP ping requests sent in a short period can quickly scan the whole network, but also increases the likelihood of being detected by Intrusion Detection Systems (IDS), which may expose the attacker. Conversely, if the scanning speed is too slow, the attacker will require a longer period to execute host discovery, thus reducing the chances of being detected. For instance, when a worm scans a network at a rate of one host every 10 seconds, it can evade all behavior-based worm detection mechanisms \cite{DBLP:conf/raid/StaffordL10}. However, in networks that employ IP shuffling, the difficulty for attackers is significantly increased, as any illegal connections initiated by the attacker must be interrupted after each IP mutation, thereby reducing both exploitation time and the accumulation of experience.

\subsubsection{Scanning Strategy}
A strong correlation exists between the addresses scanned, the current time, and historical scanning behavior. A skilled adversary may employ the following scanning strategies:
\begin{itemize} 
\item Local-preference scanning: The adversary prefers to scan adjacent IP addresses after successfully scanning an active host. 
\item Sequential scanning: The adversary probes hosts sequentially, starting from a random position within the scanning space. 
\item Uniform random scanning: The adversary randomly probes hosts, particularly after failing to discover hosts over a period of time. 
\item Hybrid scanning: The adversary randomly alternates between different scanning strategies over a period to evade detection. 
\end{itemize}

\section{LLM design}\label{sec:LLM}
\begin{figure}[!htbp]
    \centering
    \includegraphics[width=1\linewidth, trim=0 20 0 5]{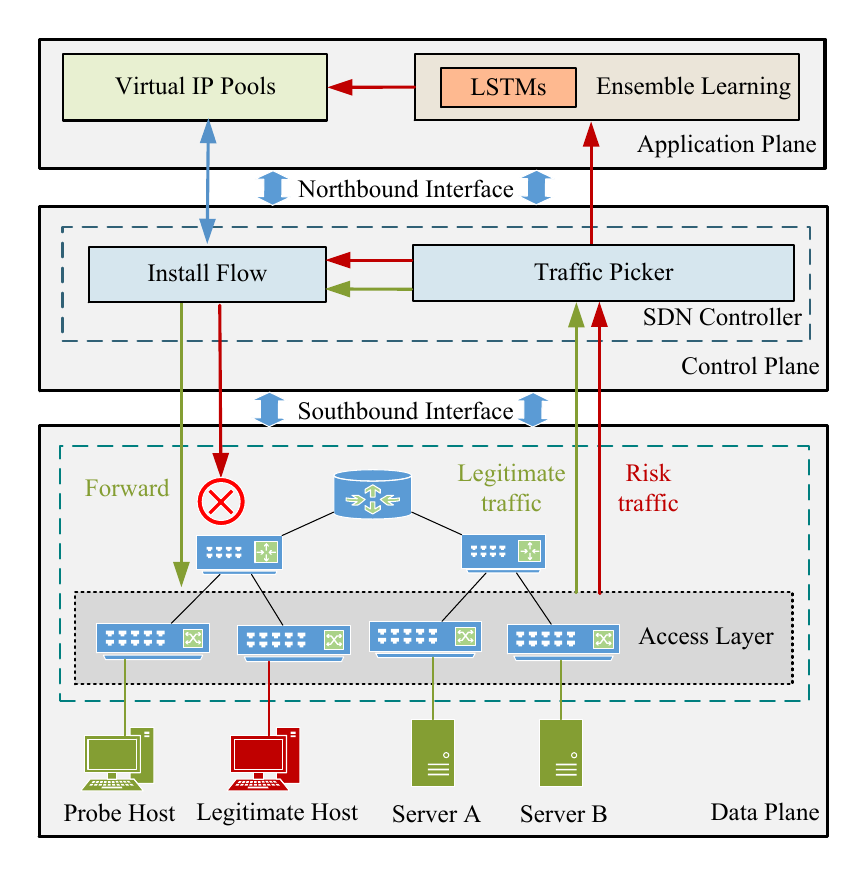}
    \caption{The architecture of LLM}
    \label{fig:arch}
\end{figure}

In this section, we introduce the architecture of LLM. LLM is built on the SDN framework to prevent internal networks from malicious reconnaissance by predicting future target addresses based on scanning behavior. When the adversary initiates suspicious scanning activities, LLM collects relevant information through a traffic picker deployed in the SDN control plane. This information is then transmitted to the application plane, where it is processed and used as input data for the LSTM prediction model. Based on the forecast result of LSTM, LLM generates corresponding virtual IPs for address allocation. Fig. \ref{fig:arch} provides an overview of the main components of LLM and their interactions.

\subsection{Software-Defined Networking}

Software-Defined Networking (SDN) is an emerging network paradigm that provides flexible and programmable infrastructure by decoupling the control and data planes. The flow table is a tuple containing actions and matches, which determines the forwarding strategies of packets in the data plane.  The SDN controller plays a central role in managing the network, allowing for dynamic configuration of flow tables and making the network more adaptive and flexible in the face of constant changes. The controller communicates to the data plane via the southbound interface, with OpenFlow being the predominant protocol. OpenFlow, which separates the data and control planes, was first proposed by McKeown in 2008 \cite{DBLP:journals/ccr/McKeownABPPRST08}. In contrast, the northbound interface, which links the application and control planes, has yet to be standardized.

SDN has become an important platform to defend against network attacks, including DDoS\cite{10705345}, scanning\cite{8406313}, and intrusion \cite{10.1007/s10586-024-04430-6}. In this work, SDN serves as the platform for efficiently deploying and managing LLM. LLM operates across both the control and application planes of the SDN architecture. The network controller continuously monitors the network and adjusts the IP addresses of hosts by modifying IP flow entries. After flow information is collected and analyzed by the LSTM in the application plane, the SDN controller performs address mutation by installing the appropriate flow entries. 

\subsection{Traffic Picker}
The access layer serves as a critical point for defenders to gather reconnaissance information about adversaries. We have deployed a traffic picker at this layer, which is responsible for monitoring and collecting any suspicious probes within the network. However, since the detection packets from attackers and legitimate user requests are not fundamentally different, it becomes difficult to identify malicious traffic. Moreover, if every packet were to be processed, network congestion would quickly ensue.

To solve the two difficulties mentioned above, LLM uses two simple mechanisms: destination filter and message queue. When a packet does not match any flow table entries, it will generate a \textit{Packet\_In} message and forward it to the SDN controller, providing metadata information, including the destination IP address. For any \textit{Packet\_In} message, traffic picker filters it based on its destination address. If the destination address belongs to an already allocated vIP or is not in the address pool, the packet will install flow table entries normally. If the destination address is an unused IP in the address pool, traffic picker will mark this probe as a risk probe. These risk probes will be used to train the LSTM model, which predicts the destination address of the future risk probe. However, processing within the SDN controller can degrade its performance. We implemented message queues to support asynchronous processing. These queues can handle millions of short messages per second, while keeping latency at the millisecond level.

\subsection{LSTM and Ensemble Learning}
Recurrent Neural Networks (RNNs) have proven effective for training on sequences of address values. Long Short-Term Memory (LSTM) networks, a variant of RNNs, are particularly well-suited for capturing long-term dependencies in data sequences, addressing the issues of vanishing and exploding gradients. LLM uses a single-layer LSTM for prediction, where predictions are mapped to 256 outputs by a fully connected layer representing all possible addresses for a standard 255.255.255.0 subnet. softmax is used as the activation function, and accordingly cross-entropy is used to calculate loss and thus predict the range of addresses an attacker is most likely to scan. We did not use a complex model because LLM wants to make real-time predictions as quickly as possible, and make changes to the flow table before the attacker establishes a connection.

\begin{figure}[!ht]
    \centering
    \includegraphics[width=1\linewidth, trim=0 15 0 15]{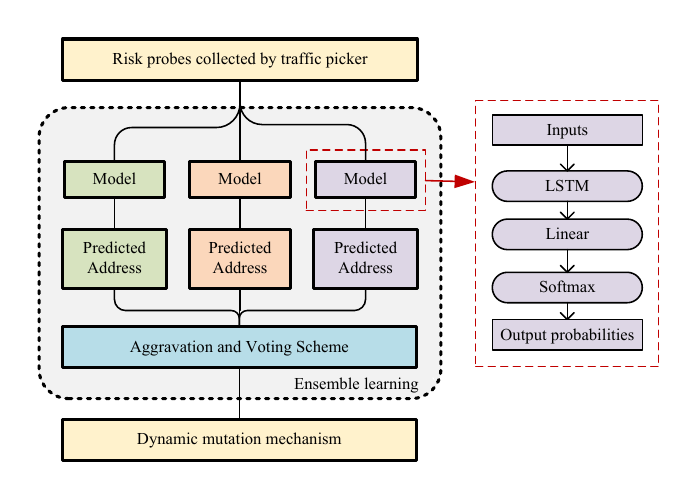}
    \caption{The framework of ensemble prediction in LLM.}
    \label{fig:ensemble}
\end{figure}

Relying solely on the predictions of a single RNN model is insufficient to enhance both the robustness and accuracy of the defense system, as attacker behavior is highly variable and contingent upon their decision-making processes. Attackers may resort to random, aimless scanning strategies to identify hosts, which are inherently unpredictable. Consequently, the LLM scheme is designed to achieve robust defense performance even under random scanning scenarios by predicting multiple scanning patterns. To this end, we leverage ensemble learning techniques. Ensemble learning combines the outputs of multiple weak models, generating a stronger prediction by integrating these results through voting mechanisms, which often yields better performance than that of any individual model. Furthermore, ensemble learning offers improved scalability, allowing the system to adapt to emerging scanning patterns by incorporating additional models. Fig. \ref{fig:ensemble} illustrates the architecture of the ensemble learning framework within the LLM scheme. Specifically, three LSTM models are employed to predict the next destination address of attackers under sequential, local preference, and random scanning behaviors. The predictions generated by these models are aggregated in a voting system, where the final predicted addresses are determined.

\section{Implementation}\label{sec:impl}
\subsection{Communication Protocols}
\begin{figure*}[!ht]
    \centering
    \includegraphics[width=1\linewidth, trim=0 20 0 20]{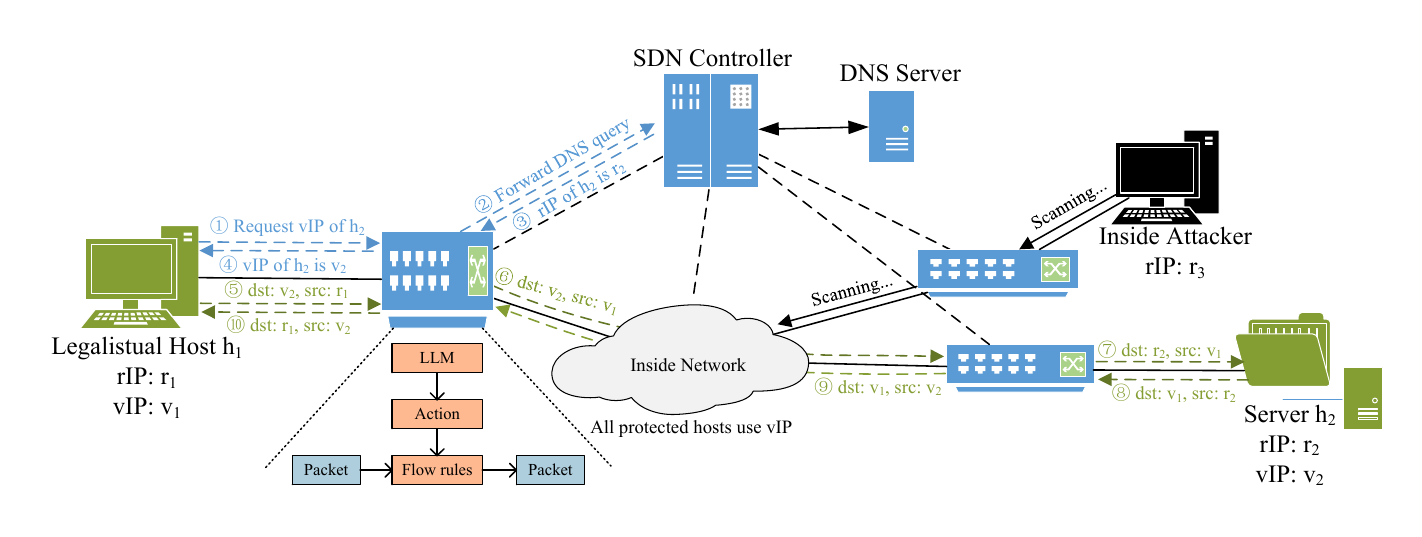}
    \caption{Communication flowchart of LLM in SDN.}
    \label{fig:fig12}
\end{figure*}

Each entity in the LLM network is blind to the real IP address (rIP) of the target host they wish to communicate with. Therefore, the source host must first obtain the virtual IP address (vIP) of the destination host in order to initiate communication. Similar to other address mutation schemes \cite{DBLP:journals/tifs/JafarianAD15,DBLP:journals/tifs/ZhangXSKG23,DBLP:conf/trustcom/SharmaK0LCM18}, LLM uses a name-to-vIP rotation mechanism, which is implemented via DNS. Fig. \ref{fig:fig12} illustrates the general flow of communication between two entities in the LLM network, facilitated through the DNS server. The communication protocol proceeds as follows:

The client sends a DNS query to obtain the vIP of the target server. After verifying the legitimacy of the client, the SDN controller responds by replacing the rIP of server with a newly generated vIP. The client then receives the vIP corresponding to the name of server. Using the vIP as the destination address, the client initiates a formal connection to the server. As the request packets pass through the access layer switches of client, the source rIP is replaced with the corresponding vIP. All subsequent data packets above the access layer use vIPs as both the source and destination addresses. When the packets reach the access layer switch of the destination, the destination vIP is replaced with the corresponding rIP. The SDN controller is responsible not only for ensuring the translation between vIP and rIP, but also for installing the necessary flow entries along the path for packets that do not have matching flows. 

\subsection{Mutation Mechanism}

\begin{figure}[!htbp]
    \centering
    \includegraphics[width=1\linewidth, trim=20 15 20 15]{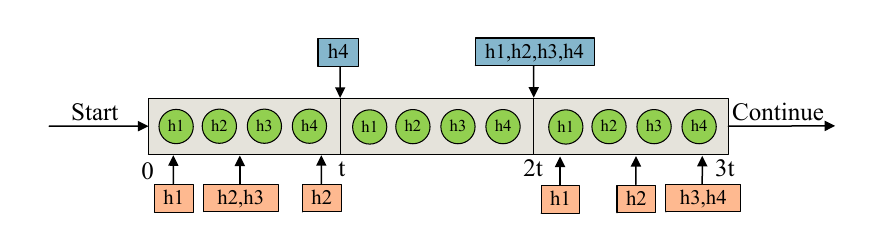}
    \caption{Mutation rotation mechanism.}
    \label{fig:mutation}
\end{figure}

In the LLM scheme, the mutation interval is dynamic, distinguishing it from previous schemes that use fixed time slots for mutation. As shown in Fig. \ref{fig:mutation}, each host in LLM has two mutation intervals: a soft mutation interval (orange box) and a hard mutation interval (blue box). The soft mutation interval is determined based on whether the currently assigned vIPs fall within the risk addresses predicted by the LSTM model. This prediction is made in real-time, characterizing the behavior of attackers.  If no scanning behavior is detected by traffic picker, the LLM scheme will execute the default hard mutation interval $t$. A hard mutation interval is implemented to minimize the impact of an attacker maintaining an established connection. After the hard mutation interval expires, the vIP of host will change, even if the vIP is not included in the predicted risk address set of the LSTM. This mechanism ensures that all hosts in the local area network undergo at least one mutation within a specified time frame.   

\section{Experimental Evaluation}\label{sec:Evaluation}
\subsection{Experiment Setup}
We utilized Mininet 2.3 \cite{mininet} to establish a network running on Ubuntu 18.04 LTS. The network consists of 3 switches and 8 hosts in a subnet with an available IP pool size of 256. LLM is implemented by Ryu 4.34 \cite{Ryu} and PyTorch 1.13.1. Ryu serves as the remote SDN controller, running on a 3.3 GHz CPU. The LSTM model is trained with PyTorch framework on Nvidia RTX 3080 GPU. The hyperparameters for the LSTM model are detailed in Table \ref{tab:table3}. There are no public and suitable network scan dataset, so we designed a simulation program to mimic attacker scanning behaviors, as described in Section \ref{sec:Models}. It is important to note that as a defense system, LLM has a complete data collection, training, and defense framework that can be deployed to any vulnerable network. The attacker uses the typical scanning tool Nmap 7.9 \cite{Nmap} for scanning. The addresses scanned by the attackers will be collected by traffic picker and recorded in CSV files as raw data. A total of 30k records are collected for each scanning strategy, with $20\%$ (6k) randomly sampled for testing, and the remaining $80\%$ (24k) records used for training. The simulated network link built using Mininet operates at a data rate of 10 Mbps, with a 0.5 ms delay on each link to reflect realistic network conditions.

\begin{table}[!ht]
\caption{Primary parameter configurations of LSTM\label{tab:table3}}
\centering
\begin{tabular}{|c||c|}
\hline Primary parameter & Setting value\\
\hline Batch size $h$ & 64\\
\hline Learning rate $h$ & 0.0001\\
\hline Epochs & 50 \\
\hline Number of hidden nodes & 100\\
\hline Activation function & Softmax\\
\hline Loss function & Cross entropy\\
\hline Class number & 256\\
\hline
\end{tabular}
\end{table}

\subsection{Defense Performance}
The hit attempts is employed as a key metric to evaluate the effectiveness of defense mechanisms. A higher number of virtual IP addresses (vIPs) scanned by the attacker indicates a lower efficacy of the Moving Target Defense (MTD) strategy. A static network configuration is utilized as the baseline for comparison. FRVM \cite{DBLP:conf/trustcom/SharmaK0LCM18} mutates IP addresses by randomly selecting unallocated IP addresses as vIPs. FRVM facilitates host mutations within a finite address pool, effectively expanding the available address space by incorporating ports as an extension of the mutation process. In contrast, Random Host Mutation (RHM) represents another IP shuffling technique, which necessitates a significantly larger IP pool to operate effectively.

\begin{figure}[!ht]
    \centering
    \includegraphics[width=0.8\linewidth, trim=0 10 0 20]{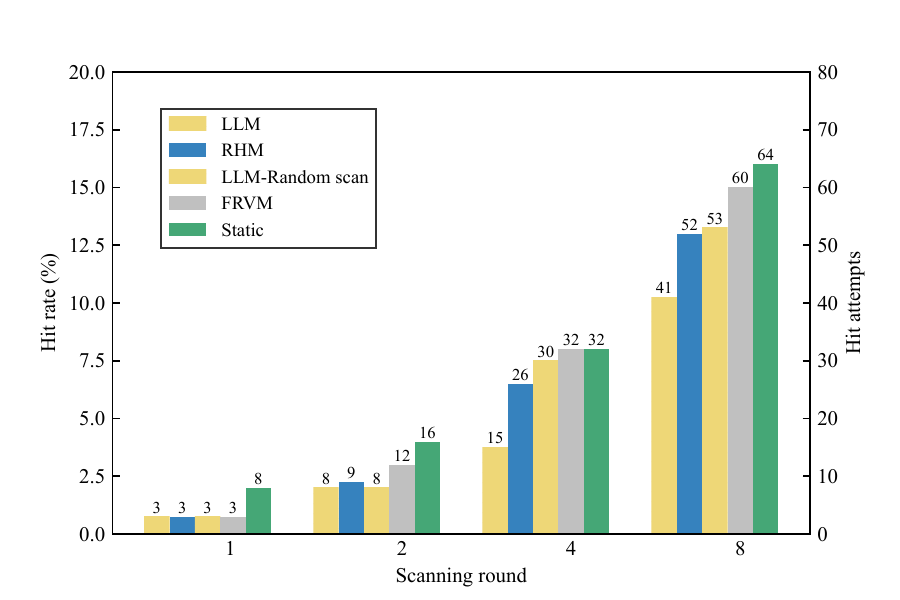}
    \caption{Defense performance comparison of four scanning strategies under network scenario with 3 switches, 8 hosts, and 256 available addresses.}
    \label{fig:defense}
\end{figure} 

We employ hybrid scanning, which is both the most effective and widely utilized attack strategy, to simulate the behavior of an attacker, thereby assessing the effectiveness of ensemble learning in LLM. Additionally, we evaluate the defensive performance of LLM against purely random scanning, where the attacker selects a new target for each scan iteration. Fig. \ref{fig:defense} presents a comparison of defense performance within a network scenario consisting of 3 switches, 8 hosts, and 256 available addresses. In comparison to static networks, the enumerated defense strategies have effectively reduced the frequency of scanning incidents.   Although FRVM extends the address space, its defensive efficacy diminishes swiftly following several scanning cycles because of the lacke of adaptivity. RHM incorporates an adaptive mechanism, primarily through hypothesis testing, which sustains its defensive capability across multiple scanning rounds. LLM markedly surpasses both aforementioned methodologies, shows the minimal number of dynamic scans over numerous cycles. Moreover, the performance of LLM is on par with RHM when subjected to random scans, attributable to our robust dynamic mutation strategy.


\subsection{Model Performance}
\begin{figure}[!ht]
\centering
\subfloat[sequential scan]{\includegraphics[width=0.45\linewidth, trim=20 20 20 25]{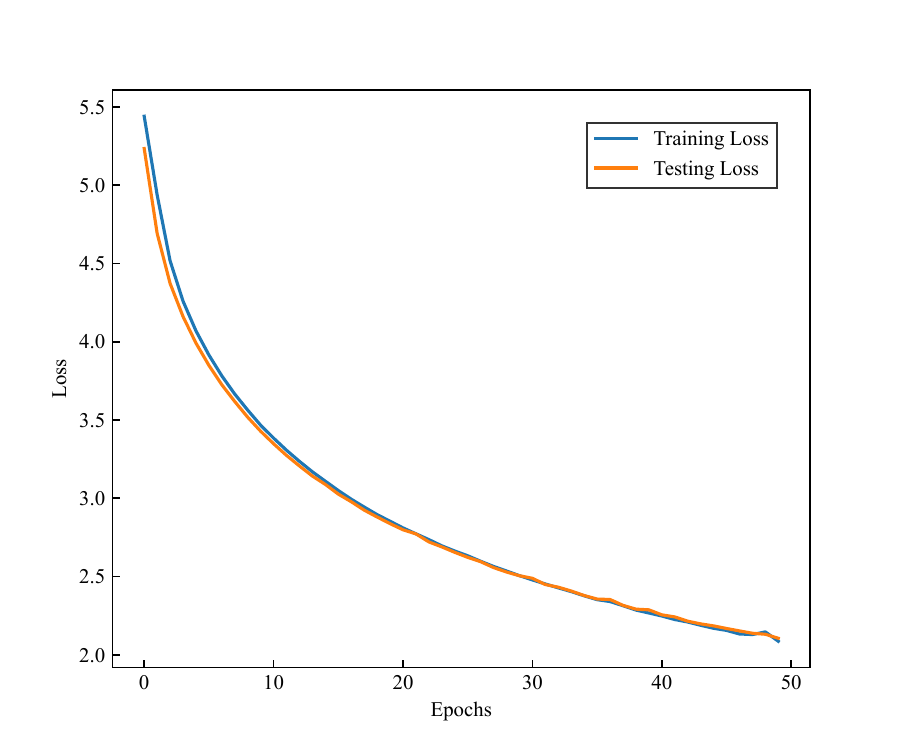}%
\hfil
\includegraphics[width=0.45\linewidth, trim=20 20 20 25]{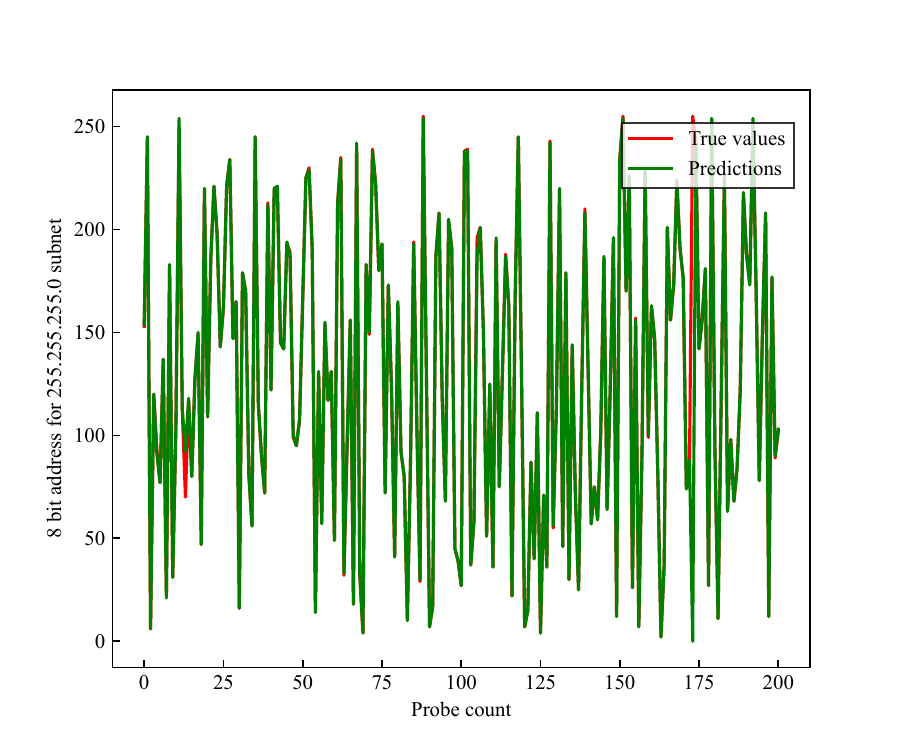}%
\label{fig_second_case}}
\hfil
\subfloat[local preference scan]{\includegraphics[width=0.45\linewidth, trim=10 20 10 40]{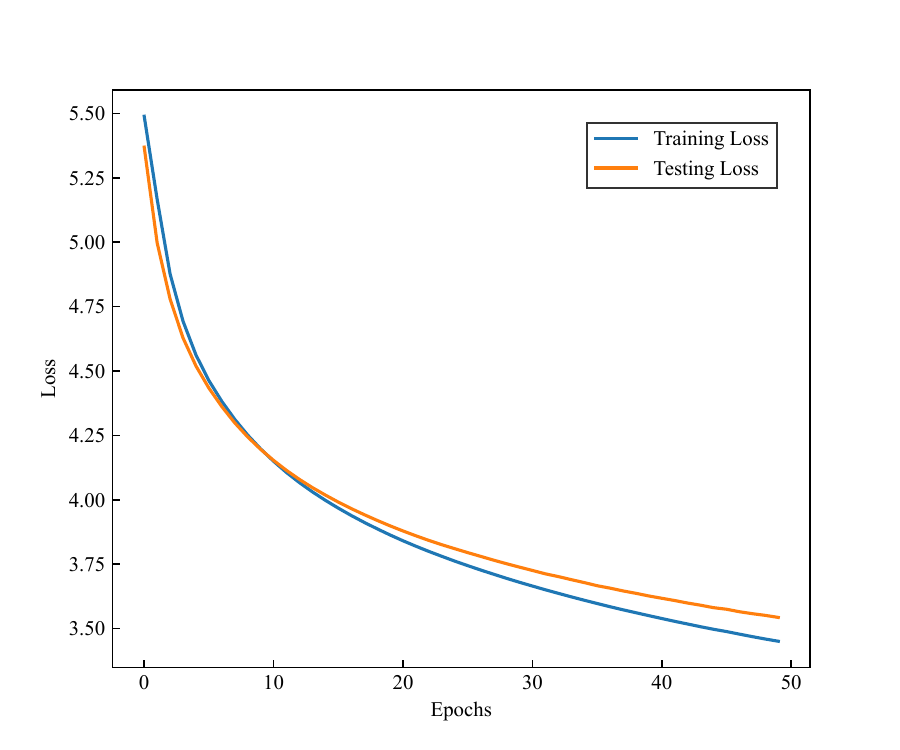}%
\hfil
\includegraphics[width=0.45\linewidth, trim=10 20 10 40]{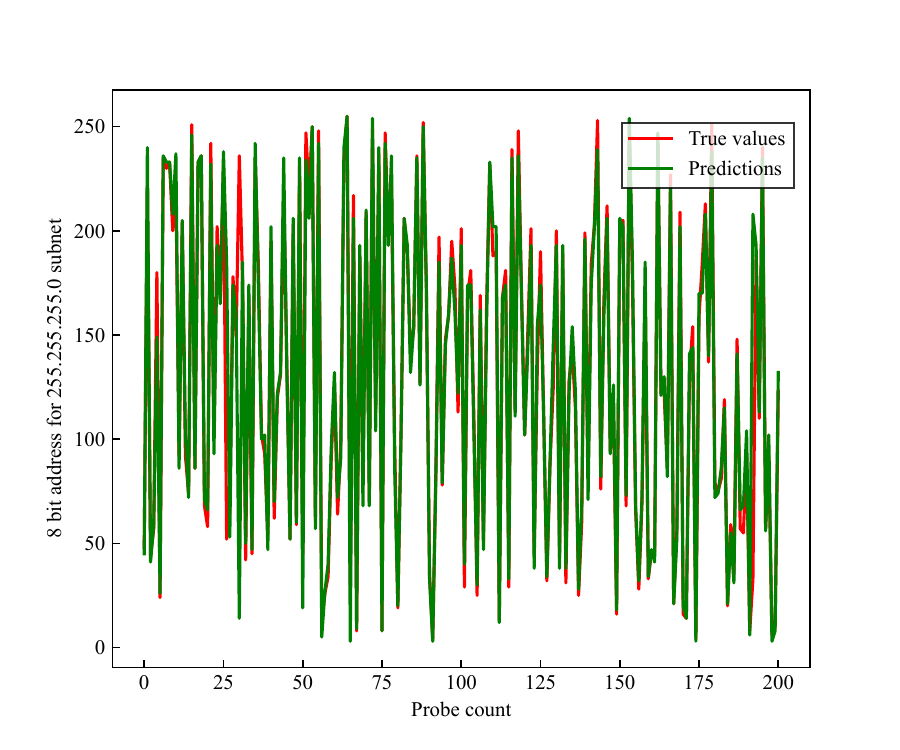}%
\label{fig_second_case2}}
\caption{Loss curve and prediction performance of LSTM in response to sequential and local preference scans (a) sequential scan. (b) local preference scan}
\label{fig_sim}
\end{figure}

In LLM, we do not care about the inference accuracy of the model, since LLM only needs to infer a fuzzy range for legitimate hosts to avoid those addresses that are most likely to be detected. Nevertheless, we validate the performance of the LSTM model in coping with address sequence prediction. Fig. \ref{fig_sim} illustrates the loss curves and prediction results of the LSTM model when coping with sequential scanning and local preference scanning. It can be intuitively seen that the LSTM model has good performance in predicting addresses. The inference time is a key metric; the faster the inference time, the faster the malicious scans that can be countered. Therefore, in Table \ref{tab:hp}, we evaluate the inference time and prediction effect under different parameters. When the number of hidden nodes was set to 100, the LSTM model achieved optimal predictive performance. The average inference time of LLM is 1.1 ms, which is significantly faster than the scanning rate of attackers in real-world scenarios (typically on the order of 100 ms). Consequently, the model demonstrates strong adaptability to both low-speed and high-speed scanning activities.

\begin{table}[!ht]
\caption{Network performance of LLM}\label{tab:hp}
\centering
\begin{tabular}{|c||c|c|c|}
\hline Hidden nodes & Infer time (ms) & RMSE & MAE\\
\hline 50 & 1.06  & 23.93 & 3.15\\
\hline 100 &1.17 & 20.03 & 2.47\\
\hline 200 & 1.11  & 22.86 & 2.97\\
\hline 400 & 1.12 & 22.27 & 2.79\\
\hline
\end{tabular}
\end{table}

\subsection{Network Performance}\label{sec:np}
To evaluate the network performance of LLM, we selected several key metrics: round-trip time (RTT), jitter, packet loss rate, and bandwidth. These metrics were measured using Iperf 3.1.3 \cite{Iperf} and the ping console tool to characterize the transmission quality of the network. We assessed the impact of dynamic mutation rate on network performance under varying scanning rates and compared the results to a static scheme as a baseline. 

\begin{figure}[!ht]
    \centering
    \includegraphics[width=0.8\linewidth, trim=0 10 0 20]{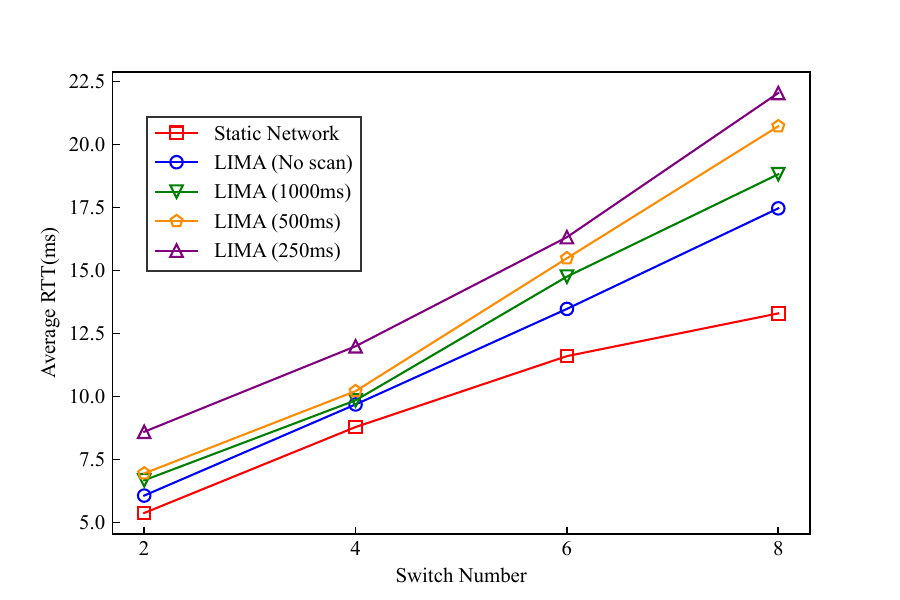}
    \caption{Network performance of RTTs.}
    \label{fig:fig5}
\end{figure}

Fig. \ref{fig:fig5} illustrates the round-trip time (RTT) of LLM at various scan rates, compared to a static network configuration. The results show that RTT increases nearly linearly with the number of switches, as additional switches introduce longer transmission paths. In the absence of scanning activity, LLM performs hard mutations at a default interval of 10 seconds, which introduces additional latency. As the scanning rate rises, the number of flow entries that need to be adjusted also increases, triggering the activation of soft mutations. This further adds to the overhead required for flow table adjustments. Overall, the increase in RTT caused by LLM is linear and can be controlled within a manageable range.

\begin{table}[!ht]
\caption{Network performance of LLM}\label{tab:np}
\centering
\begin{tabular}{|c|c||c|c|c|}
\hline Strategy & probes/s & Jitter (ms) & Loss Rate (\%) & BW (Mbps)\\
\hline
Static & 0 & 5.115 & 0.0 & 10.0\\
\hline \multirow{4}{*} {LLM}& 0 & 5.145 & 0.0 & 10.0\\
\cline{2-5}  & 1  & 5.509 & 0.0 & 9.70 \\
\cline{2-5}  & 2  & 5.127 & 0.62 & 9.70\\
\cline{2-5}  & 4  & 5.138 & 0.0 & 9.68 \\
\hline
\end{tabular}
\end{table}

We conducted comprehensive measurements of LLM using Iperf and present the results in Table \ref{tab:np}. Jitter, which represents variations in latency, increases when a new flow is installed. LLM causes only a slight increase in jitter, with the increase not exceeding $10\%$. Additionally, LLM achieves nearly zero packet loss across various scanning rates. Only one packet was lost when the scanning rate was 2 probes per second. This loss occurred because a packet temporarily missed the flow table entry update, occurring during the interval between the deletion of the old flow entry and the installation of the new one. This type of packet loss is tolerable. Furthermore, we also measured the bandwidth during scanning activities. The results show that the bandwidth of normal communication decreases during scanning, as malicious scanning behavior inevitably occupies a small portion of the available bandwidth, approximately $3\%$.

\section{Conclusion}\label{sec:Conclusion}
We present LLM, a look-ahead host address mutation strategy designed to safeguard hosts within restricted LAN environments against network reconnaissance. LLM is motivated by the insight that the target addresses of network reconnaissance correlate with both current time and historical patterns. By analyzing the scanning behaviors of attackers, we design an efficient and lightweight LSTM model to predict the next target address. The SDN controller utilizes these predictions to dynamically update flow tables, effectively avoiding IP addresses most likely to be scanned. To enhance adaptability to diverse scanning patterns, LLM integrates ensemble learning into the LSTM framework. Additionally, we introduce a dynamic mutation mechanism to ensure that all hosts undergo at least one mutation within a defined hard interval. To validate the effectiveness of our scheme, we conduct extensive experiments. The results demonstrate that LLM is highly effective and adaptive in disrupting network reconnaissance activities.

\bibliographystyle{IEEEtran}
\bibliography{LinWHWC16.bib}
\end{document}